\documentstyle[epsbox,preprint,eqsecnum,aps]{revtex}

\def\gr{\raise.3ex\hbox{$>$\kern-.75em\lower1ex\hbox{$\sim$}}}
\def\le{\raise.3ex\hbox{$<$\kern-.75em\lower1ex\hbox{$\sim$}}}
\mathchardef\Lag="724C 
\begin{document}
\draft


\preprint{US-98-02/Revised}

\def\thefootnote{\fnsymbol{footnote}}

\title{Coupling Constants Evolution \\
in  a Universal Seesaw Mass Matrix Model}

\author{Yoshio Koide\thanks{E-mail: koide@u-shizuoka-ken.ac.jp}}
\address{Department of Physics, University of Shizuoka \\
52-1 Yada, Shizuoka 422-8526, Japan}

\date{\today}

\maketitle

\begin{abstract}
Stimulated by a recent development of the universal seesaw mass 
matrix model, the evolutions of the gauge  and Yukawa coupling 
constants are investigated under the gauge symmetries
SU(3)$_c \times$SU(2)$_L\times$SU(2)$_R\times$U(1)$_Y$.
Especially, an investigation is made as to whether this evolution 
can constrain the necessary intermediate scales in these types 
of models and its viability.
\end{abstract}

\pacs{12.15.Ff, 11.10.Hi, 12.60.-i}

\narrowtext

\section{Introduction}
\label{sec:level1}


Recently, the so-called ``universal seesaw mass matrix 
model"\cite{ref1} 
has been revived\cite{ref2,ref3} as a model which 
gives a unified description of 
masses and mixings of the quarks and leptons.
The ``seesaw mechanism"  was first proposed\cite{ref4} in order to 
answer the question of why neutrino masses are so invisibly small.
Then, in order to understand that the observed quark and lepton 
masses are considerably smaller than the electroweak scale 
$\Lambda_L=\langle \phi_L^0\rangle =174$ GeV,
the mechanism was applied to the quarks\cite{ref1}. 
However, the observation of the top quark of 1994\cite{ref5} aroused
a doubt on the validity of the seesaw mechanism for the quarks
because the observed fact $m_t \sim \Lambda_L$ means that 
$M_F^{-1}m_R$ is of the order of one in the seesaw expression 
$M_f \simeq m_L M_F^{-1} m_R$.
On the contrary, it has recently been found\cite{ref2,ref3}
that the model 
can give an interpretation for the question of  why only top 
quark acquires a mass of the order of $\Lambda_L$ 
if we take an additional condition det$M_F=0$ for up-quark sector.

In the universal seesaw mass matrix model, the mass matrix for 
fermions $(f, F)$ is given by 
\begin{equation}
M = \left(\begin{array}{cc}
0 & m_L \\
m_R & M_F \\
\end{array} \right) = m_0 \left(\begin{array}{cc}
0 & Z_L \\
\kappa Z_R & \lambda Y_F \\
\end{array} \right) \ \ , 
\end{equation}
where $f_i$ (fermion sector names $f = u, d, \nu, e$; 
family numbers $i = 1, 2, 3$) 
denote quarks and leptons, $F_i$ denote hypothetical heavy fermions 
$F = U, D, N$ and $E$ correspondingly to $f = u, d, \nu$ and $e$, 
and they belong to $f_L = (2,1)$, $f_R = (1,2)$, $F_L = (1,1)$ 
and $F_R = (1,1)$ of SU(2)$_L \times $SU(2)$_R$.
The matrices $Z_L$, $Z_R$ and $Y_F$ are those 
of the order of one.
The $3\times 3$ matrices $m_L$ ($\sim m_0=\Lambda_L$) and 
$m_R$ ($\sim \kappa m_0=\Lambda_R$) are symmetry breaking 
mass terms of SU(2)$_L$ and SU(2)$_R$, respectively, and 
those have common structures independently of 
the fermion sector names $f$. 
Only $M_F$ ($\sim \lambda m_0=\Lambda_S$) has a structure 
dependent on the sector name $f$. 
For the case $\lambda \gg \kappa \gg 1$, 
the mass matrix (1.1) leads to the well-known seesaw expression
\begin{equation}
 M_f \simeq m_L M_F^{-1} m_R  \ .
\end{equation}

In contrast  to the case (1.2), for the case with the additional 
condition 
\begin{equation}
{\rm det}M_F =0 \ ,
\end{equation}
on the up-quark sector ($F=U$), one of the heavy fermions $F_i$
(say, $F_3$) cannot acquire a mass of the order of 
$\Lambda_S\equiv \lambda m_0$, 
so that the seesaw mechanism does not work for the third fermion.
Therefore, the mass generation at each energy scale is as follows:
First, at the energy scale $\mu=\Lambda_S$, the heavy fermions $F$,
except for $U_3$, acquire the masses of the order of $\Lambda_S$.
Second, at the energy scale $\mu=\Lambda_R$, the SU(2)$_R$ 
symmetry is broken, and the fermion $u_{R3}$ generates a mass
term of the order of $\Lambda_R$ by pairing with $U_{L3}$.
Finally, at $\mu=\Lambda_L$, the SU(2)$_L$ symmetry is broken,
and the fermion $u_{L3}$ generates a mass term of the order 
$\Lambda_L$ by pairing with $U_{R3}$. 
The other fermions $f$ acquire the well-known seesaw masses
(1.2).
The scenario is summarized in Table \ref{tablemf}.
We regard the fermion pair $(u_{L3}, U_{R3})$ as the top-quark 
state. 
Thus, we can understand why only top quark $t$ acquires  
the mass $m_t \sim O(m_L)$\cite{ref2,ref3}.

On the other hand, for the neutrino mass generation, 
at present, we have the following two scenarios 
as summarized in Table \ref{tablemnu}.
One (Scenario A) is a trivial extension of 
the present model:
we introduce a further large energy scale 
$\Lambda_{\nu S}$ in addition to $\Lambda_S$, and
we assume that $M_F \sim \Lambda_S$ ($F=U,D,E$), while 
$M_N \sim \Lambda_{\nu S}$ ($\Lambda_{\nu S}\gg \Lambda_S$).
Another scenario (Scenario B)\cite{ref6} is one without introducing 
such an additional energy scale.
The neutral heavy leptons are singlets of 
SU(2)$_L\times$SU(2)$_R$ and they do not have
U(1)-charge. 
Therefore, it is likely that they acquire 
Majorana masses $M_M$ together with the Dirac 
masses $M_D\equiv M_N$ at $\mu=\Lambda_S$.
For example, we assume $M_M=M_D$\cite{ref7}.
Then, the neutrino mass matrix for the conventional 
light neutrinos is given by $M_\nu=m_L M_N^{-1} m_L^T$,
so that the masses $m_\nu$ are given with the order of 
\begin{equation}
m_\nu \sim \frac{\Lambda_L^2}{\Lambda_S}=
\frac{1}{\kappa}\frac{\Lambda_L\Lambda_R}{\Lambda_S} \ .
\end{equation}
In order to explain the smallness of $m_\nu$, 
the model requires that the scale $\Lambda_R$ must be extremely 
larger than $\Lambda_L$ (for example, 
$\kappa\equiv \Lambda_R/\Lambda_L\sim 10^9$\cite{ref7}).
This scenario seems to be very attractive from the 
theoretical point of view, because we can 
explain the mass hierarchy of the quarks and leptons by 
the three energy scales $\Lambda_L$, $\Lambda_R$ and 
$\Lambda_S$ only.
On the other hand, in the scenario A, there is no 
constraint on the value of $\kappa$ (however, the value
must be larger than $\sim 10$ because of no observation of 
the right-handed weak bosons $W_R$ at present), so that
the model allows a case with a lower value of $\Lambda_R$.
Since we can expect abundant new physics effects for 
the case of $\kappa\sim 10$\cite{ref8}, the case is also attractive
from the phenomenological point of view. 

One of the purposes of the present paper is to see whether
a study of the evolutions of the gauge coupling constants 
of SU(3)$_c\times$SU(2)$_L\times$SU(2)$_R\times$U(1)$_Y$ and
of the Yukawa coupling constants in the universal seesaw 
mass matrix model can give any hint on the value of 
the intermediate energy scale $\Lambda_R$ or not.
For example,  Shafi and Wetterich\cite{ref9} and 
Rajpoot\cite{ref10} 
have considered an O(10) model and an SO(10) model, 
respectively, with the symmetry breakings 
SO(10)$\rightarrow$ 
SU(3)$_c \times$SU(2)$_L\times$SU(2)$_R\times$U(1)$_Y$
at $\mu=\Lambda_{GUT}$ and SU(2)$_R \rightarrow$U(1)$_R$ at 
$\mu=\Lambda_R$, and they have demonstrated  that the model with
$\Lambda_{GUT}\sim 10^{19}$ GeV and $\Lambda_R\sim 10^9$ GeV is consistent 
with the low energy phenomenology.
The value $\Lambda_R\sim 10^9$ GeV is favorable to the scenario B
for neutrino masses. 
However, in the present model, since there are many new fermions $F$
above the intermediate energy scale $\Lambda_S$, their conclusion  cannot be
applied to the present seesaw mass matrix model straightforwardly.

On the other hand, a phenomenological study of the universal seesaw mass 
matrix model for the quark masses and the 
Cabibbo-Kobayashi-Maskawa (CKM)\cite{refckm} matrix parameters
has successfully been given by Fusaoka and the author\cite{ref2}.
In order to give explicit numerical predictions, they have used
some working hypotheses that I will use here as well.

\noindent
(i) The matrices $Z_L$ and $Z_R$, which are universal
for quarks and leptons, have the same 
structure:
\begin{equation}
Z_L = Z_R \equiv Z = {\rm diag} (z_1, z_2, z_3) \ \ , 
\end{equation}
with $z_1^2 + z_2^2 + z_3^2 = 1$, 
where, for convenience, we have taken a basis on which 
the matrix $Z$ is diagonal. 

\noindent
(ii) The matrices $Y_F$, which have structures 
dependent on the fermion sector $f=u,d,\nu,e$, take
a simple form [(unit matrix)+(a rank one matrix)]:
\begin{equation}
Y_f = {\bf 1} + 3 b_f X \ \ . 
\end{equation}
(iii) The rank one matrix $X$ is  given by
a democratic form
\begin{equation}
X = \frac{1}{3}\left(\begin{array}{ccc}
1 & 1 & 1 \\
1 & 1 & 1 \\
1 & 1 & 1 \\
\end{array} \right) \  , 
\end{equation}
on the family-basis where the matrix $Z$ is diagonal.

\noindent
(iv) In order to fix the parameters $z_i$, we 
tentatively take $b_e = 0$ for the charged lepton sector,
so that the parameters $z_i$ are given by
\begin{equation}
\frac{z_1}{\sqrt{m_e}} = \frac{z_2}{\sqrt{m_\mu}} = 
\frac{z_3}{\sqrt{m_\tau}} = \frac{1}{\sqrt{m_e + m_\mu + m_\tau}} \ \ . 
\end{equation}
By taking $b_u=-1/3$ (then det$M_U=0$), they have obtained 
the following top-quark mass enhancement without the suppression 
factor $\kappa/\lambda$
\begin{equation}
m_t \simeq \frac{1}{\sqrt{3}} m_0 \ ,
\end{equation}
together with the successful relation $m_u/m_c \simeq 
3m_e/4m_\mu$.
Furthermore, by taking $b_d=-e^{i\beta_d}$ ($\beta_d=18^\circ$),
they have succeeded in giving the reasonable values of the CKM matrix
parameters together with the reasonable values of the quark mass ratios 
(not only $m_i^u/m_j^u$, $m_i^d/m_j^d$,  but also 
$m_i^u/m_j^d$) with keeping the value of the parameter 
$(m_0 \kappa/\lambda)_f$ in $(m_0 \kappa/\lambda)_u=
(m_0\kappa/\lambda)_d$. 
However, in order to fit the quark mass values (not 
the ratios) to the observed quark mass values at 
$\mu=m_Z$, they have taken the parameter $(m_0\kappa/\lambda)_f$  
as
\begin{equation}
R(m_Z)\equiv \left(\frac{(m_0\kappa/\lambda)_u}{
(m_0\kappa/\lambda)_e}\right)_{\mu=m_Z}\simeq 3 \ . 
\end{equation}
It seems to be natural to consider that all Yukawa 
coupling constants become equal between quarks and 
leptons at a large  energy scale $\Lambda_{YU}$.
Therefore, another one of the purposes of the present paper is 
to see whether such a factor 3 can be understood by the difference 
of the evolutions of the Yukawa coupling constants 
between quarks and  leptons from the energy scale 
$\mu=\Lambda_{YU}$ to $\mu=m_Z$.

In Sec.~II and Sec.~III, we investigate  evolution of the 
gauge  and Yukawa coupling constants,
respectively, under the gauge symmetries 
SU(3)$_c\times$SU(2)$_L\times$SU(2)$_R\times$U(1)$_Y$
at one-loop. 
We will conclude that it is possible to find the energy 
scale $\Lambda_{YU}$ at which $R(\mu)$ takes $R=1$ only 
for a model with a value $\kappa <10^2$. 
Although in Sec.~II and Sec.~III we consider the case that 
the symmetries 
SU(3)$_c\times$SU(2)$_L\times$SU(2)$_R\times$U(1)$_Y$ are
unbroken for the region $\mu >\Lambda_S$, in Sec.~IV, 
we investigate a case that the symmetries 
SU(3)$_c\times$U(1)$_Y$ are embedded into
the Pati-Salam symmetry\cite{refpati} SU(4)$_{PS}$ at 
$\mu > \Lambda_S$, so that we consider the case of 
SU(4)$_{PS}\times$SU(2)$_L\times$SU(2)$_R$ in the region
$\Lambda_S <\mu\leq \Lambda_{GUT}$. 
We will find that the model predicts 
$\Lambda_R \simeq 5\times 10^{12}$ GeV, 
$\Lambda_S \simeq 3\times 10^{14}$ GeV, 
and $\Lambda_{GUT} \simeq 6\times 10^{17}$ GeV. 
Finally, Sec.~V is devoted to the conclusions and remarks.
We will find that there is no model which satisfies
$\Lambda_{YU}=\Lambda_{GUT}$. 



\vspace{.2in}

\section{Evolution of the gauge coupling constants}
\label{sec:level2}


The gauge symmetries 
SU(3)$_c\times$SU(2)$_L\times$SU(2)$_R\times$U(1)$_Y$
are broken into the gauge symmetries 
SU(3)$_c\times$SU(2)$_L\times$U(1)$_{Y'}$ 
at $\mu=\Lambda_R$.
The electric charge operator $Q$ 
\begin{equation}
Q=I_3^L +I_3^R + \frac{1}{2} Y \ ,
\end{equation}
at $\mu>\Lambda_R$ 
is changed into
\begin{equation}
Q=I_3^L +\frac{1}{2} Y' \ ,
\end{equation}
in the region $\Lambda_L <\mu\leq \Lambda_R$.
Hereafter, we call the regions $\Lambda_L <\mu\leq \Lambda_R$, 
$\Lambda_R <\mu\leq \Lambda_S$ and $\Lambda_S <\mu\leq \Lambda_X$ 
($\Lambda_X\equiv \Lambda_{GUT}$ or $\Lambda_X\equiv \Lambda_{YU}$)
Regions I, II and III, respectively.

The evolutions of the gauge coupling constants $g_i$ at
one-loop are given by the equations
\begin{equation}
\frac{d}{dt} \alpha_i(\mu) = -\frac{1}{2\pi}
 b_i \alpha_i^2(\mu) \  , 
\end{equation}
where $\alpha_i\equiv g_i^2/4\pi$, $t=\ln \mu$ and the coefficients 
$b_i$ are given in Table \ref{tableb}.
(Note that the heavy fermions $F_L$ and $F_R$ except for $U_{L3}$
and $U_{R3}$ are decoupled for $\mu\leq\Lambda_S$ and the 
fermions $u_{R3}$ and $U_{L3}$ are decoupled for 
$\mu\leq\Lambda_R$.) 
The boundary conditions at $\mu=\Lambda_L$ and 
$\mu=\Lambda_R$ are as follows:
\begin{equation}
\alpha^{-1}_{em}(\Lambda_L) = \alpha_L^{-1}(\Lambda_L)+
\frac{5}{3}\alpha_1^{\prime -1}(\Lambda_L) \ ,
\end{equation}
and
\begin{equation}
\frac{5}{3}\alpha^{\prime -1}_{1}(\Lambda_R) = 
\alpha_R^{-1}(\Lambda_R)+
\frac{2}{3}\alpha_1^{-1}(\Lambda_R) \ ,
\end{equation}
respectively, correspondingly to (2.2) and (2.1), where
the normalizations of the U(1)$_{Y'}$ and U(1)$_Y$ gauge 
coupling constants have been taken as they satisfy 
$\alpha'_1=\alpha_L=\alpha_3$ in the SU(5) grand-unification
limit and $\alpha_1=\alpha_L=\alpha_R=\alpha_3$ in the 
SO(10) grand-unification limit, respectively.
For convenience, we use the initial values at $\mu=m_Z$ 
instead of those at $\mu=\Lambda_L$ in the region I:
$\alpha'_1(m_Z)=0.01683$, $\alpha_L(m_Z)=0.03349$ 
and $\alpha_3(m_Z)=0.118$. 
The values of $\alpha'_1$ and  $\alpha_L$ have been derived 
from \cite{refhollik} $\alpha_{em}(m_Z)=(128.89\pm0.09)^{-1}$ 
and $\sin^2\theta_W=0.23165\pm 0.00024$. 
The value of $\alpha_3$ has been quoted from Ref.\cite{refpdg}.
We illustrate a typical case with $\Lambda_R=10^5$ GeV and 
$\alpha_R(\Lambda_R)=1/4\pi$ in Fig.~1.

In the numerical study, we have taken the value of the parameter
$\Lambda_R/\Lambda_S\equiv \kappa/\lambda$ as 
\begin{equation}
\kappa/\lambda = \Lambda_R/\Lambda_S = 0.02  \ , 
\end{equation}
which has been obtained form the observed value of the ratio
 $m_c/m_t$ in Ref.\cite{ref2}.
Although the value (2.6) has been obtained on the model with
the specific matrix forms (1.5) - (1.7), the order of the 
value (2.6) will be valid for any other seesaw model with 
det$M_U=0$  because in such a model 
the value of $\kappa/\lambda$ is given by the order of $m_b/m_t$.

As seen in Fig.~1, the U(1) coupling constant $\alpha_1(\mu)$ 
becomes rapidly strong in the region III ($\mu>\Lambda_S$) 
because the heavy fermions $F$ become massless in the 
region III.
We consider that the unification energy scale $\Lambda_{YU}$ 
of the Yukawa coupling constants must be lower than an 
energy scale $\Lambda_1^\infty$ at which $\alpha_1(\mu)$ 
becomes infinity.
This condition will impose a strong restriction on the 
possible $\Lambda_{YU}$-search as we discuss in the next
section.
Of course, in the grand unification scenario, the U(1) 
symmetry will be embedded into a grand unification 
symmetry $G$ before the U(1) coupling constant bursts.
Such a case will be discussed in Sec.~IV.

\vspace{.2in}
\section{Evolution of ${\text{\lowercase{\it y}}}_L 
{\text{\lowercase{\it y}}}_R/{\text{\lowercase{\it y}}}_S$}
\label{sec:level3}

The $3\times 3$ matrices $m_L$, $m_R$ and $M_F$ are given
in terms of the vacuums expectation values 
$v_L=\sqrt{2}\langle\phi_L^0\rangle$, 
$v_R=\sqrt{2}\langle\phi_R^0\rangle$ and 
$v_S=\langle\Phi\rangle$, and the matrices $Z$ and $Y_F$ 
defined by (1.5) - (1.7) as follows:
\begin{equation}
m_L^f = \frac{1}{\sqrt{2}} y_L^f v_L Z\ , \ \ 
m_R^f = \frac{1}{\sqrt{2}} y_R^f v_R Z\ , \ \ 
M_F = y_S^f v_S Y_F \ . 
\end{equation}
The evolution of the Yukawa coupling constants are given by
\begin{equation}
\frac{d}{dt} \ln (y_L^f Z) = \frac{1}{16\pi^2} (T_L^f - G_L^f + H_L^f)  \ , 
\end{equation}
\begin{equation}
\frac{d}{dt} \ln (y_R^f Z) = \frac{1}{16\pi^2} (T_R^f - G_R^f + H_R^f)  \ , 
\end{equation}
\begin{equation}
\frac{d}{dt} \ln (y_S^f Y_F) = \frac{1}{16\pi^2} (T_S^f - G_S^f + H_S^f)  \ , 
\end{equation}
where $T^f$, $G^f$ and $H^f$ denote contributions from fermion-loop 
corrections, vertex corrections due to the gauge bosons and vertex 
corrections due to the Higgs boson, respectively.

What is of great interest to us is to see whether the evolutions can 
explain the value $R(m_Z) \simeq 3$ or not, i.e.,
our interest exists not in the hierarchy among $m_e$, $m_\mu$ 
and $m_\tau$, but in the hierarchy among up-quark, down-quark, 
charged lepton and neutrino sectors.
Therefore,  we neglect the scale-dependency of the matrix $Z$,
because we can regard the value of $z_3$ as $z_3\simeq 1$ 
from Eq.~(1.8). 
We also neglect the scale-dependency of the matrix $Y_F$ 
because the matrices $Y_F$ are expressed as
\begin{equation}
Y_F = {\rm diag}(1,1,1+3b_f) \ ,
\end{equation}
on the basis on which the matrix $Y_F$ is diagonal, and 
we find that the forms $Y_E= $diag$(1,1,1)$ and 
$Y_U= $diag$(1,1,0)$ are scale-invariant and  
$Y_D= $diag$(1,1,1-3e^{i\beta_d})$ is almost scale-invariant.
For convenience, we approximately still use the evolution equation
of $y_S Y_F$ at $\mu\leq \Lambda_S$ and that of $y_R Z$ 
at $\mu\leq\Lambda_R$.
Then, the ratio $R(\mu)$ defined by Eq.~(1.10) can be 
expressed in terms of the Yukawa coupling constants $y_L$, 
$y_R$ and $y_S$ as follow:
\begin{equation}
R(\mu) \equiv \frac{y_L^u (\mu) y_R^u (\mu)/y_S^u(\mu)}{
y_L^e (\mu) y_R^e (\mu)/y_S^e(\mu)} \ . 
\end{equation}
The evolution of the ratio $R(\mu)$ is approximately given by 
\begin{equation}
\frac{d}{dt} \ln R(\mu) \simeq -\frac{1}{16\pi^2} (G - H) \ , 
\end{equation}
where
\begin{eqnarray}
G= & (G_L^u+G_R^u-G_S^u)-(G_L^e+G_R^e-G_S^e) \ , \nonumber \\
H= & (H_L^u+H_R^u-H_S^u)-(H_L^e+H_R^e-H_S^e) \ . \nonumber \\
\end{eqnarray}
The $G$- and $H$-terms are given in Table \ref{tabley}.
Since in the present model, $|y_L^u|^2\simeq |y_L^d|^2$, 
$|y_L^e|^2\simeq |y_L^\nu|^2$, and so on, 
differently from other models where 
$|y_L^d/y_L^u|\simeq m_b/m_t$ and $|y_L^\nu/y_L^e|\simeq 
m_\nu/m_\tau$,
we can neglect the $H_L$- and $H_R$-terms in Eq.~(3.8).
When we also neglect the $H_S$-terms, the ratio $R(\mu)$
is approximately evaluated as follows:
\begin{equation}
\frac{R(\mu)}{R(m_Z)} = \left(1 + \frac{b_3^{I}}{2\pi} \alpha_3 (m_Z) 
\ln \frac{\mu}{m_Z} \right)^{-4/b_3^{I}}
\left(1 + \frac{b_1^I}{2\pi} \alpha'_1 (m_Z) 
\ln \frac{\mu}{m_Z} \right)^{7/10b_1^I}  \ , 
\end{equation}
\begin{equation}
\frac{R(\mu)}{R(\Lambda_R)} = \left(1 + \frac{b_3^{II}}{2\pi} \alpha_3 
(\Lambda_R) \ln \frac{\mu}{\Lambda_R} \right)^{-4/b_3^{II}} 
\left(1 + \frac{b_1^{II}}{2\pi} \alpha_1 
(\Lambda_R) \ln \frac{\mu}{\Lambda_R} \right)^{1/b_1^{II}} \ , 
\end{equation}
\begin{equation}
\frac{R(\mu)}{R(\Lambda_S)} = \left(1 + \frac{b_3^{III}}{2\pi} \alpha_3 
(\Lambda_S) \ln \frac{\mu}{\Lambda_S} \right)^{-4/b_3^{III}} 
\left(1 + \frac{b_1^{III}}{2\pi} \alpha_1 
(\Lambda_R) \ln \frac{\mu}{\Lambda_S} \right)^{1/b_1^{III}} \ , 
\end{equation}
for the regions I, II and III, respectively. 
By using (3.9) - (3.11), 
we can obtain the energy scale $\mu = \Lambda_{YU}$ 
at which the ratio $R(\mu)$ takes $R(\Lambda_{YU})=1$.

In Fig.~2, we illustrate the behavior of $\Lambda_{YU}$
for a given value of $\Lambda_R$.
For reference, we also illustrate the behavior of 
$\Lambda_1^\infty$, at which $\alpha_1^{-1}(\mu)$ takes
 $\alpha_1^{-1}(\Lambda_1^\infty)=0$.
The value of $\Lambda_{YU}$ must be lower than the
value of $\Lambda_1^\infty$.
Therefore, as seen in Fig.~2, if we adhere to the 
constraint $\alpha_R(\Lambda_R)=\alpha_L(\Lambda_R)$, we 
must abandon a model with a higher $\kappa$ value 
$(\kappa >10^2)$.
Only a model with $\kappa\sim 10$ is acceptable.
However, if we admit a strong coupling of the right-handed 
weak bosons at $\mu=\Lambda_R$, for example, 
$\alpha_R(\Lambda_R)\geq 1/4\pi$, a model with a higher 
$\kappa$ value also becomes acceptable.

Of course, from a similar study, we can find that the
evolution of $R_{u/d}(\mu)\equiv (y_L^u y_R^u /y_S^u)/
(y_L^d y_R^d/ y_S^d)$ still keeps $R_{u/d}(m_Z)\simeq 1$.
Therefore, the parametrization $(m_0 \kappa/\lambda)_u=
(m_0\kappa/\lambda)_d$ in Ref.\cite{ref2} is justified.

\vspace{.2in}
\section{Evolution of the Pati-Salam color}
\label{sec:level4}

In order to avoid the burst of the U(1) gauge coupling 
constant, we consider that the U(1)$_Y\times$SU(3)$_c$ 
symmetries are embedded into the Pati-Salam SU(4) symmetry
\cite{refpati} above $\mu=\Lambda_S$.
In other words, the SU(4)$_{PS}$ gauge symmetry is broken 
into SU(3)$_c\times$U(1)$_Y$ at $\mu=\Lambda_S$.
Indeed, the structures of the heavy fermion mass matrices 
$M_F$ are flavor-dependent.
The fermions $f$ and $F$ belong to $f_L=(2,1,4)$, 
$f_R=(1,2,4)$, $F_L=(1,1,4)$ and $F_R=(1,1,4)$ of 
SU(2)$_\times$SU(2)$_R\times$SU(4)$_{PS}$ at
$\mu \geq \Lambda_S$.

In the region III ($\Lambda_S<\mu\leq \Lambda_X$),
$\alpha_L(\mu)$ and $\alpha_R(\mu)$ are 
evolved with the coefficients 
$b_L^{III}$ and $b_R^{III}$ given in Table \ref{tableb},
but $\alpha_1(\mu)$ and $\alpha_3(\mu)$ 
are replaced with $\alpha_4(\mu)$ which is evolved 
with the coefficient
\begin{equation}
b_4^{III}=20/3 \ , 
\end{equation}
where the boundary condition at $\mu=\Lambda_S$ is 
\begin{equation}
\alpha_1(\Lambda_S)=\alpha_3(\Lambda_S)=
\alpha_4(\Lambda_S)\ .
\end{equation}
Since $\alpha_1(\Lambda_S)$ and $\alpha_3(\Lambda_S)$
are given by
\begin{equation}
\alpha_1^{-1}(\Lambda_S)=\frac{5}{2}\left[ 
\alpha_1^{-1}(\Lambda_L)+\frac{b_1^{I}}{2\pi}
\ln\frac{\Lambda_R}{\Lambda_L}\right] 
-\frac{3}{2}\left[ \alpha_L^{-1}(\Lambda_L)+\frac{b_L^{I}}{2\pi}
\ln\frac{\Lambda_R}{\Lambda_L}\right]
+\frac{b_1^{II}}{2\pi}\ln\frac{\Lambda_S}{\Lambda_R}\ ,
\end{equation}
\begin{equation}
\alpha_3^{-1}(\Lambda_S)=\alpha_3^{-1}(\Lambda_L)+
\frac{b_3^{I}}{2\pi}\ln\frac{\Lambda_R}{\Lambda_L}+
\frac{b_3^{II}}{2\pi}\ln\frac{\Lambda_S}{\Lambda_R} \ ,
\end{equation}
respectively, the values of $\Lambda_R$ and $\Lambda_S$ 
are fixed at
\begin{equation}
\Lambda_R=5.46\times 10^{12} \ {\rm GeV}\ , \ \ \ \ 
\Lambda_S=2.73\times 10^{14} \ {\rm GeV}\ ,
\end{equation}
under the conditions (2.6) and (4.2).
The unification scale $\Lambda_{GUT}$ is also fixed at
\begin{equation}
\Lambda_{GUT}=5.84\times 10^{17} \ {\rm GeV}\ , 
\end{equation}
by the condition
\begin{equation}
\alpha_L(\Lambda_{GUT})=\alpha_R(\Lambda_{GUT})=
\alpha_4(\Lambda_{GUT}) \ ,
\end{equation}
for example, for the embedding into SO(10)\cite{ref8,ref9}.
In Fig.~3, we illustrate the behaviors of $\alpha_i^{-1}(\mu)$.
Roughly speaking, the value $\Lambda_R\sim 10^{12}$ is 
favorable to the scenario B for the neutrino mass generation.

On the other hand, the evolution of $R(\mu)$ defined by
(3.6) is almost constant at $\mu \geq \Lambda_S$, i.e.,
$R(\Lambda_S)\simeq R(\Lambda_{YU})$, because there 
is no difference between quarks and leptons in the 
region III ($\Lambda_S<\mu \leq \Lambda_{YU}$).
Therefore,  we obtain
\begin{equation}
R(m_Z)/R(\Lambda_{YU})\simeq R(m_Z)/R(\Lambda_S)=2.3\ ,
\end{equation}
and we fail to obtain our desirable relation
$R(m_Z)/R(\Lambda_{YU})\simeq 3$.
If we adhere the unification of the gauge symmetries
SU(3)$_c\times$SU(2)$_L\times$SU(2)$_R\times$U(1)$_Y$
into a Pati-Salam type unification $G$, we must abandon
the idea that the discrepancy $R(m_Z)\simeq 3$ between
quarks and leptons in the model given in Ref.\cite{ref2}
comes from difference of evolutions between quarks and
leptons, or we must consider that the magnitudes of
the Yukawa coupling constants are different between
quarks and leptons from the beginning at $\mu=\Lambda_{GUT}$.

\vspace{.2in}

\section{Concluding remarks}
\label{sec:level4}

In conclusion, we have investigated the evolution of the 
universal seesaw mass matrix model under the gauge symmetries 
SU(3)$_c\times$SU(2)$_L\times$SU(2)$_R\times$U(1)$_Y$. 
The symmetries can be embedded into a unification 
symmetry of the Pati-Salam type at 
$\Lambda_{GUT}=5.84\times 10^{17}$ GeV.
Then, the value $\Lambda_R=5.46\times 10^{12}$ GeV is 
favorable to the scenario B for the neutrino mass generation.
However, we cannot explain the discrepancy $R(m_Z)\simeq 3$ between 
quarks and leptons by the evolution of $R(\mu)$ starting from 
$R(\Lambda_{GUT})=1$. 

On the other hand, if we abandon the grand unification 
scenario, the model has a possibility that the value
$R(m_Z)\simeq 3$ can be understood by the evolution of the 
Yukawa coupling constants.
As seen in Fig.~2, we require 
$\alpha_L(\Lambda_R)=\alpha_R(\Lambda_R)$, 
the value $\Lambda_R$ for the case which gives $R(m_Z)\simeq 3$
must be $\Lambda_R \leq 10^4$ GeV.
If we accept a model with a strong SU(2)$_R$ force at 
$\mu=\Lambda_R$, for example, 
$\alpha_R(\Lambda_R)=1/4\pi$,
the region $\Lambda_R\leq 10^{18}$ GeV  also becomes allowed.
We consider that the model with $\kappa\sim 10$ is likely.
Although this case rules out the scenario B for neutrinos,
phenomenologically we can expect an abundance of new 
physics effects \cite{ref8}, $t'$ production, FCNC effects, 
and so on, in the near future colliders.

Our numerical results have been obtained by the 
constraint $\Lambda_R/\Lambda_S=0.02$ which has 
come from the observed ratio of $m_c/m_t$ under
the special model with (1.5) - (1.7).
Since the ratio $\Lambda_R/\Lambda_S$ 
is fixed by the ratio $m_c/m_t$ (or 
$m_b/m_t$) as far as a universal seesaw model 
with det$M_U=0$ are concerned, 
the value of the ratio $\Lambda_R/\Lambda_S$
is, in general, of the order of $10^{-2}$.
Therefore, our conclusions will be unchanged 
as far as the orders are concerned.

In the present paper, we have not discussed a 
SUSY version of the present model, although
the case is attractive from the point of view of
the grand unification.
In such a SUSY version, since the coefficient 
$8\alpha_3$ in $G$-terms in (3.8) 
[also in Table V]
is changed for $(16/3)\alpha_3$, the case push 
the energy scale $\Lambda_{YU}$ to an unlikely 
ultra-high energy scale ($>10^{23}$ GeV).
If we want to adopt a SUSY version of the present 
model, we must abandon the idea of the unification
of the Yukawa coupling constants.

\vspace*{.2in}
\vglue.4in

\acknowledgments

The author would like to thank M.~Tanimoto, T.~Matsuoka 
and N.~Okamura for their helpful comments, 
especially, on the evolution equations.
This work was supported by the Grand-in-Aid for Scientific
Research, Ministry of Education, Science and Culture,
Japan (No.~08640386).

\vglue.2in

%

\begin{table}
\caption{Fermion mass generation scenario \label{tablemf}}

\begin{tabular}{|c|c|c|}\hline
Energy scale &  $d$- \& $e$-sectors &
$u$-sector ($i\neq 3$) \\ \hline
At $\mu=\Lambda_S\sim \lambda m_0$ & $m(F_L, F_R) \sim \Lambda_S$  
& $m(U_{Li}, U_{Ri}) \sim \Lambda_S$ \\ \hline
At $\mu=\Lambda_R\sim \kappa m_0$ &  & $m(u_{R3}, U_{L3}) 
\sim \Lambda_R$  \\ \hline 
At $\mu=\Lambda_L\sim m_0$ & 
& $m(u_{L3}, U_{R3})\sim \Lambda_L$  \\[.2in] 
 & $m(f_L, f_R)\sim \displaystyle\frac{\Lambda_L \Lambda_R}{\Lambda_S}$  & 
$m(u_{Li}, u_{Ri}) \sim \displaystyle\frac{\Lambda_L \Lambda_R}{\Lambda_S}$ 
 \\ \hline
\end{tabular}
\end{table}

\begin{table}
\caption{Neutrino mass generation scenarios:
$N_\pm =(N_L \pm N_R^c)/\protect\sqrt{2}$ \label{tablemnu}}

\begin{tabular}{|c|c|c|}\hline
Energy scale &  Scenario A & Scenario B  \\ \hline
At $\mu=\Lambda_{\nu S}$ & $m(N_L, N_R) \sim \Lambda_{\nu S}$  
& \\ \hline
At $\mu=\Lambda_S$ &  & $m(N_+, N_+^c) \sim \Lambda_S$ \\ \hline
At $\mu=\Lambda_R$ &  & $m(\nu_R, N_-)\sim \Lambda_R$  \\ \hline 
At $\mu=\Lambda_L$ & $m(\nu_L, \nu_R)\sim 
\displaystyle\frac{\Lambda_L \Lambda_R}{\Lambda_{\nu S}}$ &
$m(\nu_L, \nu_L^c) \sim \displaystyle\frac{\Lambda_L^2}{\Lambda_S}$  \\ \hline
\end{tabular}

\end{table}

\begin{table}
\caption{ Quantum numbers of the fermions $f$ and $F$
and Higgs scalars $\phi_L$, $\phi_R$ and $\Phi$. \label{tableqn}}

$$
\begin{array}{|c|ccc|c|ccc|} \hline
  & I_3^L & I_3^R & Y &  & I_3^L & I_3^R & Y \\ \hline
u_L & +\frac{1}{2} & 0 & \frac{1}{3} & u_R & 0 & +\frac{1}{2} & \frac{1}{3} \\
d_L & -\frac{1}{2} & 0 & \frac{1}{3} & d_R & 0 & -\frac{1}{2} 
& \frac{1}{3} \\ \hline
\nu_L & +\frac{1}{2} & 0 & -1 & \nu_R & 0 & +\frac{1}{2} & -1 \\
e_L & -\frac{1}{2} & 0 & -1 & e_R & 0 & -\frac{1}{2} & -1 \\ \hline
U_L & 0 & 0 & \frac{4}{3} & U_R & 0 & 0 & \frac{4}{3} \\
D_L & 0 & 0 & -\frac{2}{3} & D_R & 0 & 0 & -\frac{2}{3} \\ \hline
N_L & 0 & 0 & 0 & N_R & 0 & 0 & 0 \\ 
E_L & 0 & 0 & -2 & E_R & 0 & 0 & -2 \\ \hline
\phi_L^+ & +\frac{1}{2} & 0 & 1 & \phi_R^+ & 0 & +\frac{1}{2} & 1 \\ 
\phi_L^0 & -\frac{1}{2} & 0 & 1 & \phi_R^0 & 0 & -\frac{1}{2} & 1 \\ 
\hline
\Phi & 0 & 0 & 0 &  &  &  &    \\ \hline
\end{array}
$$
\end{table}

\begin{table}
\caption{Coefficients in the evolution equations of 
gauge coupling constants.\label{tableb}}

$$
\begin{array}{|c|c|c|c|}\hline
      & \Lambda_L <\mu\leq\Lambda_R & 
 \Lambda_R <\mu\leq\Lambda_S
& \Lambda_S <\mu\leq\Lambda_X \\ \hline
{\rm SU(3)}_c & b^I_3=7 & b^{II}_3=19/3 & b^{III}_3=3 \\ \hline
{\rm SU(2)}_L & b^I_L=19/6 & b^{II}_L= 19/6 & b^{III}_L=19/6 \\ \hline
{\rm SU(2)}_R &  & b^{II}_R=19/6 & b^{III}_R=19/6 \\ 
\cline{1-1}\cline{3-4}
{\rm U(1)}_Y & {b^I_1}=-41/10 & b_1^{II}=-43/6 & b_1^{III}=-41/2 \\ \hline
\end{array}
$$
\end{table}


\begin{table}
\caption{$G$- and $H$-terms in the evolution equation  
(3.8). $H_L^e$, $H_R^e$ and $H_S^e$ are given by the 
replacements $u\rightarrow e$ and $d\rightarrow \nu$ 
in $H_L^u$, $H_R^u$ and $H_S^u$, respectively.
\label{tabley}}

$$
\begin{array}{|c|c|c|c|}\hline
      & I \ (\Lambda_L <\mu\leq\Lambda_R) & 
II \ (\Lambda_R <\mu\leq\Lambda_S)
& III \ (\Lambda_S <\mu\leq\Lambda_{YU}) \\ \hline
G= & 4\pi\left(8\alpha_3 -\frac{7}{5}\alpha'_1\right) & 
\multicolumn{2}{c|}{
4\pi\left(8\alpha_3 -2\alpha_1\right)} \\ \hline
H_L^u = & \multicolumn{3}{c|}{\frac{3}{2}\left(|y_L^u|^2-
|y_L^d|^2\right) } \\ \hline
H_R^u= &  0   & \multicolumn{2}{c|}{
\frac{3}{2}\left(|y_R^u|^2-|y_R^d|^2\right) } \\ \hline
H_S^u= & \multicolumn{2}{c|}{0} & 3|y_S^u|^2 \\ \hline
\end{array}
$$
\end{table}


\begin{figure}
\begin{center}
\epsfile{file=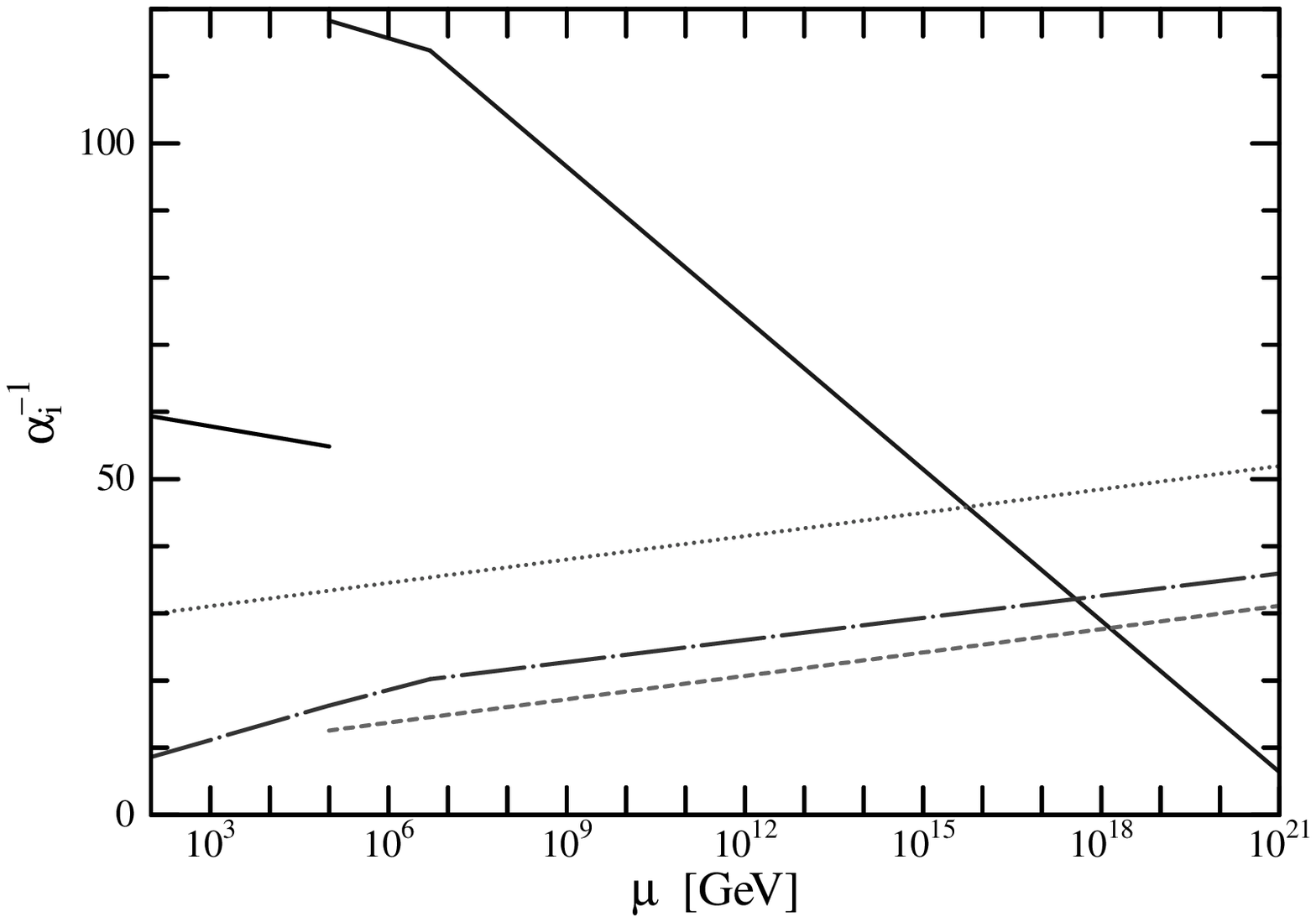,scale=0.7}
\end{center}
\caption{
Behaviors of $\alpha_3^{-1}(\mu)$ (dotted chain
line), 
$\alpha_L^{-1}(\mu)$ (dotted line), 
$\alpha_R^{-1}(\mu)$ (broken line) and 
$\alpha_1^{-1}(\mu)$ [$\alpha_1^{\prime -1}(\mu)$] (solid line) 
in the case with 
the input values $\Lambda_R=10^5$ GeV and 
$\alpha_R(\Lambda_R)=1/4\pi$.
}
\end{figure}


\vspace{1cm}

\begin{figure}
\begin{center}
\epsfile{file=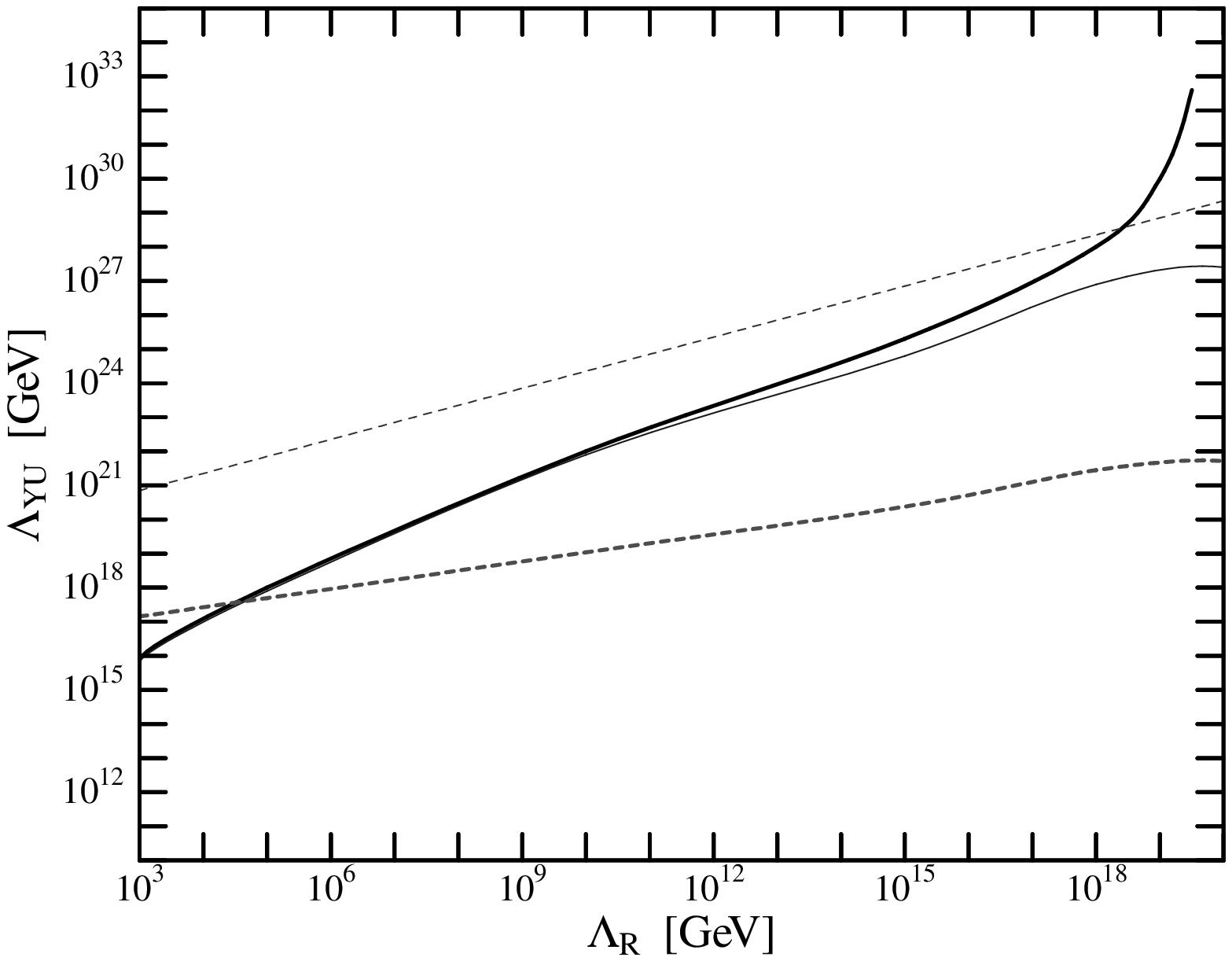,scale=0.65}
\end{center}
\caption{
Behavior of $\Lambda_{YU}$ versus $\Lambda_R$.
The bold and thin solid lines denote the cases of 
the input values $\alpha_R(\Lambda_R)= \alpha_L(\Lambda_R)$
and $\alpha_R(\Lambda_R)=1/4\pi$, respectively.
For reference, the behaviors of 
$\Lambda_1^\infty$ for the cases of 
the input values $\alpha_R(\Lambda_R)= \alpha_L(\Lambda_R)$
(bold broken line) and $\alpha_R(\Lambda_R)=1/4\pi$
(thin broken line) are illustrated.
The physical value of $\Lambda_{YU}$ must be 
$\Lambda_{YU} < \Lambda_1^\infty$.
}
\end{figure}

\vspace{1cm}

\begin{figure}
\begin{center}
\epsfile{file=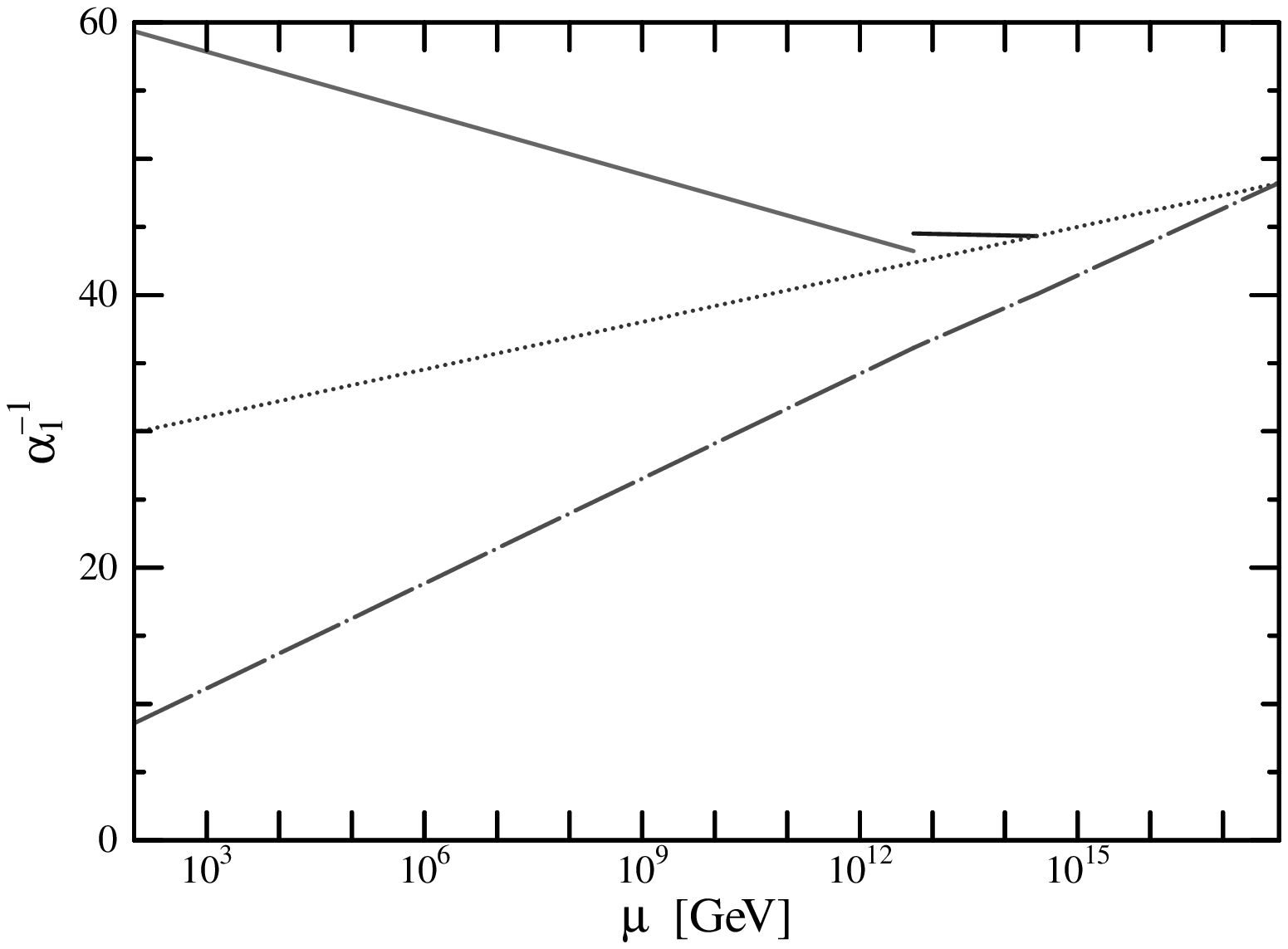,scale=0.65}
\end{center}
\caption{
Behaviors of $\alpha_3^{-1}(\mu)$ 
[$\alpha_4^{-1}(\mu)$ for $\Lambda_S<\mu\leq\Lambda_{GUT}$] 
(dotted chain line), 
$\alpha_L^{-1}(\mu)$ [$=\alpha_R^{-1}(\mu)$ for 
$\mu>\Lambda_R$] (dotted line) and 
$\alpha_1^{-1}(\mu)$ [$\alpha_1^{\prime -1}(\mu)$
for $\Lambda<\mu\leq \Lambda_R$] 
(solid line) in the case of the  
Pati-Salam type unification,
where $\Lambda_R=5.46\times 10^{12}$ GeV,
$\Lambda_S=2.37\times 10^{14}$ GeV and 
$\Lambda_{GUT}=5.84\times 10^{17}$ GeV.
}
\end{figure}


\begin{references}
%
\bibitem{ref1} Z.~G.~Berezhiani, Phys.~Lett.~{\bf 129B}, 99 (1983);
Phys.~Lett.~{\bf 150B}, 177 (1985);
D.~Chang and R.~N.~Mohapatra, Phys.~Rev.~Lett.~{\bf 58},1600 (1987); 
A.~Davidson and K.~C.~Wali, Phys.~Rev.~Lett.~{\bf 59}, 393 (1987);
S.~Rajpoot, Mod.~Phys.~Lett. {\bf A2}, 307 (1987); 
Phys.~Lett.~{\bf 191B}, 122 (1987); Phys.~Rev.~{\bf D36}, 1479 (1987);
K.~B.~Babu and R.~N.~Mohapatra, Phys.~Rev.~Lett.~{\bf 62}, 1079 (1989); 
Phys.~Rev. {\bf D41}, 1286 (1990); 
S.~Ranfone, Phys.~Rev.~{\bf D42}, 3819 (1990); 
A.~Davidson, S.~Ranfone and K.~C.~Wali, 
Phys.~Rev.~{\bf D41}, 208 (1990); 
I.~Sogami and T.~Shinohara, Prog.~Theor.~Phys.~{\bf 66}, 1031 (1991);
Phys.~Rev. {\bf D47}, 2905 (1993); 
Z.~G.~Berezhiani and R.~Rattazzi, Phys.~Lett.~{\bf B279}, 124 (1992);
P.~Cho, Phys.~Rev. {\bf D48}, 5331 (1994); 
A.~Davidson, L.~Michel, M.~L,~Sage and  K.~C.~Wali, 
Phys.~Rev.~{\bf D49}, 1378 (1994); 
W.~A.~Ponce, A.~Zepeda and R.~G.~Lozano, 
Phys.~Rev.~{\bf D49}, 4954 (1994).
%
\bibitem{ref2} Y.~Koide and H.~Fusaoka, Z.~Phys. {\bf C71}, 459 (1996); 
Prog.~Theor.~Phys. {\bf 97}, 459 (1997).
%
\bibitem{ref3} T.~Morozumi, T.~Satou, M.~N.~Rebelo and 
M.~Tanimoto, Phys.~Lett. 
{\bf B410}, 233 (1997).
%
\bibitem{ref4} M.~Gell-Mann, P.~Rammond and R.~Slansky, in {\it Supergravity}, 
edited by P.~van Nieuwenhuizen and D.~Z.~Freedman (North-Holland, 
1979); 
T.~Yanagida, in {\it Proc.~Workshop of the Unified Theory and 
Baryon Number in the Universe}, edited by A.~Sawada and A.~Sugamoto 
(KEK, 1979); 
R.~Mohapatra and G.~Senjanovic, Phys.~Rev.~Lett.~{\bf 44}, 912 (1980).
%
\bibitem{ref5}  CDF collaboration, F.~Abe $et$ $al$., Phys.~Rev.~Lett. 
{\bf 73}, 225 (1994).
%
%
\bibitem{ref6} Z.~G.~Berezhiani, in Ref.[1]; 
A.~Davidson and K.~C.~Wali, in Ref.[1];
S.~Rajpoot, in Ref.[1];
A.~Davidson, S.~Ranfone and K.~C.~Wali, in Ref.[1];
W.~A.~Ponce, A.~Zepeda and R.~G.~Lozano, in Ref.[1].
%
\bibitem{ref7} Y.~Koide, hep-ph/9707505 (1997), to be published in 
Phys.~Rev. {\bf D}, No.9 (1998). 
%
\bibitem{ref8} Y.~Koide, Phys.~Rev. {\bf D56}, 2656 (1997).
%
\bibitem{ref9} Q.~Shafi and C.~Wetterich, Phys.~Lett. {\bf 85B}, 
52 (1979).
%
\bibitem{ref10} S.~Rjpoot, Phys.~Rev. {\bf D22}, 2244 (1980).
%
\bibitem{refckm} N.~Cabibbo, Phys.~Rev.~Lett.~{\bf 10}, 531 (1996); 
M.~Kobayashi and T.~Maskawa, Prog.~Theor.~Phys.~{\bf 49}, 652 (1973).
%
\bibitem{refpati} J.~C.~Pati and A.~Salam, Phys.~Rev.~Lett. 
{\bf 31}, 661 (1973); {\bf 36}, 1229 (1976); 
Phys.~Lett. {\bf 58B}, 333 (1975); 
Phys.~Rev. {\bf D8}, 1240 (1973); {\bf D10}, 275 (1974).
%
%
\bibitem{refhollik} W.~Hollik, 
University of Karlsruhe Report No.~KA-TP-19-1996,
1996 (unpublished); See also Z.~Hioki, Acta Phys.~Pol. 
{\bf B27}, 1569 (1996).
%
\bibitem{refpdg} I.~Hinchliffe, Particle data group, 
Phys.~Rev. {\bf D54}, 77 (1996).
%
%
%
%
\end{references}
\end{document}